# 19 GHz Single-Ended-to-Balanced Modified Ladder-Lattice Filters Realized Using Periodically Polarized AlScN BAW Resonators

Merrilyn M. A. Fiagbenu, *Student Member, IEEE*, Shun Yao, *Graduate Student Member, IEEE*, Siddhant Sahoo, Mojtaba Hodjat-Shamami, *Member, IEEE*, Daeho Kim, *Member, IEEE*, Craig Moe, *Member, IEEE*, Pinal Patel, Ramakrishna Vetury, and Roy H. Olsson, III, *Senior Member, IEEE*

***Abstract*—We present a single-ended to balanced modified ladder-lattice filter topology, realized with periodically polarized piezoelectric (P3F) AlScN resonators, for single-ended to differential filtering applications. This filter topology eliminates the need for baluns or other passive components and can interface, for instance, directly between differential antennas and single-ended power and low noise amplifiers. Two filter designs operating at 19 GHz were implemented. The first design exhibited a minimum insertion loss of 1.3 dB, a 6.26% fractional 3-dB bandwidth, an average out-of-band rejection of 30 dB and occupied a 348 µm x 476 µm footprint, while the second higher-selectivity filter achieved 1.58 dB of minimum insertion loss, 5.11% fractional 3-dB bandwidth, 33 dB of average out-of-band rejection, and occupied a 392 µm x 476 µm footprint – Making them highly competitive for wireless communications filters. The working principle of the modified ladder-lattice topology is discussed, the design of the filters are presented, and the linearity of the filters is explored through intermodulation and power handling measurements.**

## I. Introduction

SINCE 2001, aluminum nitride (AlN) bulk acoustic wave (BAW) resonators have been utilized in duplexers in mobile radio systems due to their large acoustic velocity, compact size, low acoustic loss, low dielectric loss, and CMOS process compatibility, effectively replacing bulky ceramic duplexers [1-5]. An example of such an AlN duplexer is the first-generation Avago Duplexer, which was composed of AlN film bulk acoustic wave resonators (FBARs) [1]. While the 7% maximum electromechanical coupling coefficient of AlN has been sufficient to meet the bandwidth requirements (~3%) of many communication bands, it is still insufficient in meeting the demands of larger bandwidth technologies such as wideband active electronically scanned array antennas (AESAs) and innovations such as Artificial Intelligence that require multi-frequency operation or faster data processing speed and transfer rate and hence higher frequency and larger bandwidth [1-3], [6-8].

To realize acoustic resonators with higher electromechanical coupling coefficients ($k_t^2$) and high operating frequencies, researchers have explored approaches such as substitutional alloying of AlN with scandium (Sc) or adopted piezoelectric materials with higher intrinsic $k_t^2$ such as lithium niobate ($LiNbO_3$) – trading higher electromechanical coupling for lower quality factor Q (since Sc softens AlN and lowers its Q), or sacrificing CMOS-compatibility (due to lithium contamination of standard CMOS processes) [1-3], [6-7], [9-21]. To this end, Izhar et al. reported 17.4 GHz periodically polarized piezoelectric (P3F) AlScN resonators with $k_t^2$ of 11.8%, $Q_s$ ($Q_p$) of 58.4 (236.6), and a figure of merit (FoM) of 6.9 (27.9) defined as $k_t^2Q_s$ ($k_t^2Q_p$) [13]. Barrera et al. demonstrated a 19.8 GHz resonator realized in P3F $LiNbO_3$ with $k_t^2$ of 44%, $Q_s$ of 40, and $k_t^2Q_s$ FoM of 17.6 [14]. And Schaffer et al. presented overmoded AlN bulk acoustic resonators operating at 55.7 GHz with $k_t^2$ of 2.2%, $Q_s$ ($Q_p$) of 93 (89), and FoM of 2.0 (2.0) [15].

Research in this space has produced acoustic filters such as an 11.9 GHz ladder filter in P3F AlScN films by Fiagbenu et al., a 22.1 GHz ladder filter by Barrera et al., and a 23.8 GHz ladder filter by Cho et al. [22-24]. The filter by Fiagbenu et al. achieved 1.5 dB minimum insertion loss (ILmin), 6.6% 3-dB fractional bandwidth (FBW), and 26 dB out-of-band (OoB) rejection [22]. The filter by Barrera et al. demonstrated 1.62 dB minimum insertion-loss, 19.8% FBW, and 12.5 dB of out-of-band rejection [23]. Finally, Cho et al.'s filter achieved 1.52 dB of minimum IL, 19.4% FBW, and 12.1 dB of OoB rejection [24]. While state-of-the-art, many of these acoustic RF filters fail to compete with their lower frequency BAW counterparts

This work was supported by the Defense Advanced Research Project Agency (DARPA) COmpact Front-end Filters at the ElEment-level (COFFEE) Program. All views or opinions presented, including any findings, are those of the authors and should not be interpreted as representing the official views or policies of the Department of Defense or of the U.S. Government. (Corresponding author: Roy H. Olsson, III).

Merrilyn M. A. Fiagbenu, Shun Yao, and Roy H. Olsson, III, are with the Department of Electrical and Systems Engineering, University of Pennsylvania, Philadelphia, PA 19104 USA (e-mail: rolsson@seas.upenn.edu).

Siddhant Sahoo, Mojtaba Hodjat-Shamami, Daeho Kim, Craig Moe, Pinal Patel, and Ramakrishna Vetury are with Akoustis Technologies Inc., Huntersville, NC 28078 USA.

[13-14], [22-26]. They employ conventional ladder topologies that fundamentally trade out-of-band rejection for insertion loss. However, this declining trend in filter performances with increasing frequency is to be expected, given the lower FoMs ($k_t^2Q$ products) associated with operating acoustic resonators at higher frequencies [7], [13-26]. Such FoM degradation can be attributed to reductions in the electromechanical coupling coefficient of resonators operating in overtone modes, a decrease in the quality factor of BAW resonators through Akhiezer damping as their resonant frequencies increase, declining film quality such as reduced crystallinity with decreasing film thickness, as well as increased acoustic and electrical losses as the thickness of piezoelectric and electrode layers are decreased to accommodate higher frequency operation [1], [7], [13-21], [27]. Additional causes include the reduction of resonator areas while increasing resonance frequency to maintain constant resonator impedance for impedance matching to external systems, which leads to higher anchor loss for FBAR resonators [13], [16-17], [20], [27].

This paper presents a first-time demonstration of two modified ladder lattice (mLL) filters with 18.8 GHz and 19 GHz center frequencies realized in 3-layer P3F AlScN. The first filter (design type-1) has a minimum insertion loss of 1.3 dB, 6.26% FBW, and an average OoB rejection of 30 dB, while the second more-selective filter (design type-2) has an $IL_{min}$ of 1.58 dB, a 5.11% FBW, and an average OoB rejection of 33 dB. These filters employ a three-port modified ladder-lattice topology that makes adequate use of the lower quality-factors and limited impedance contrast of the resonators in the process design kit (PDK). To achieve this performance, the authors exploited the higher electromechanical coupling coefficients of AlScN films (greater than 8% in this case), a P3F resonator structure, the high selectivity of ladder filters, and the high out-of-band rejection of lattice filters [3], [6], [9], [13], [16-21], [28-30]. The P3F structure compensates for the increase in BAW capacitance with increasing frequency by increasing the effective piezoelectric thickness, thereby stabilizing the resonators' impedances and facilitating impedance matching to RF ports/systems [13], [16-17], [19-20]. Low insertion loss is accomplished by minimizing the number of filter stages.

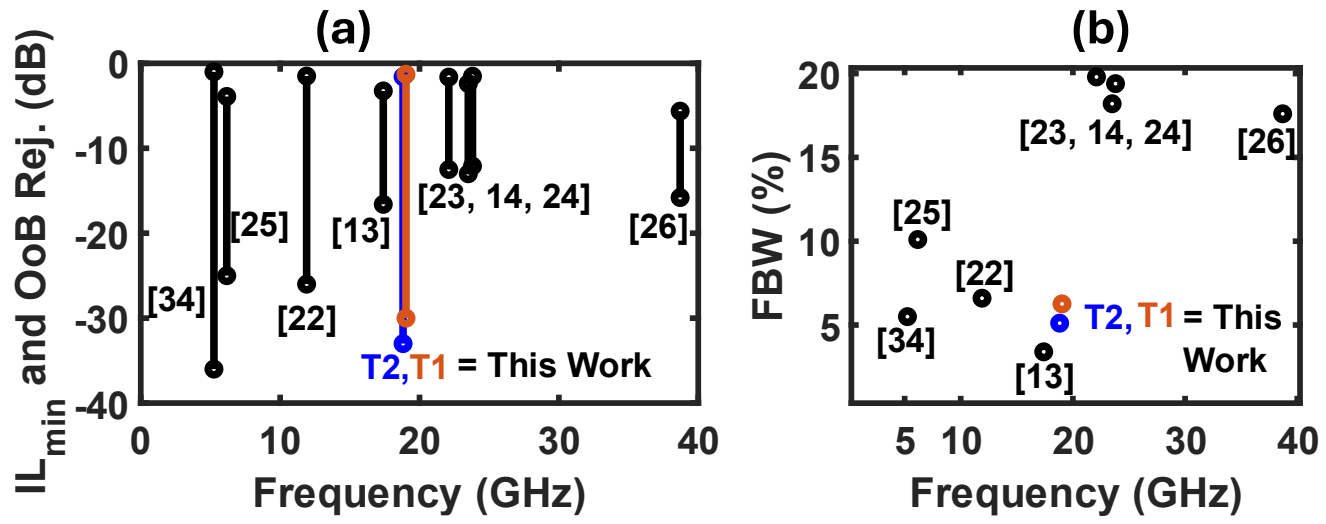

Fig. 1. A comparison of the filters in this work to those of other filters in the literature. Insertion loss (IL) and out-of-band rejection (OoB) are compared in (a), while fractional bandwidth (FBW) is compared in (b).

By eliminating the use of external components such as capacitors, inductors and baluns, the filter topology reported here presents a compact solution for single-ended to differential filtering or duplexer applications. For instance, it can interface directly between single-ended power amplifiers (PAs) or low noise amplifiers (LNAs) and differential antennas (such as standard dipoles) or between differential digital-to-analog and analog-to-digital converters and single-ended PAs and LNAs in some RF transceiver paths [29], [31-33]. Fig. 1 shows a comparison of the filters' minimum insertion losses ($IL_{min}$), out-of-band (OoB) rejection, and relative 3-dB fractional bandwidths (FBW) to those of state-of-the-art RF filters above 5 GHz [13-14], [22-26], [34].

## II. Working Principle

A conventional ladder-lattice filter design employs balanced ladder and lattice unit segments often with interchangeable locations [29-30], [34]. However, the modified ladder lattice (mLL) filter topology contains a terminal single-ended ladder segment that can serve as an input or output port, and a modified lattice segment that performs single-ended to balanced/differential conversion. Fig. 2 shows the schematics of the two implementations of the mLL filter to be discussed in this paper. While design type-1 consists of a single ladder and modified lattice unit, design type-2 contains two ladder units and one modified lattice unit, with the return port of the ladder unit that occurs after the lattice unit connecting to one of the balanced ports rather than to ground. Due to the simplicity of its design, filter type-1 will be used to explain the working mechanism of the mLL filter topology. However, filter type-2 functions in an identical manner, but with an additional filtering stage.

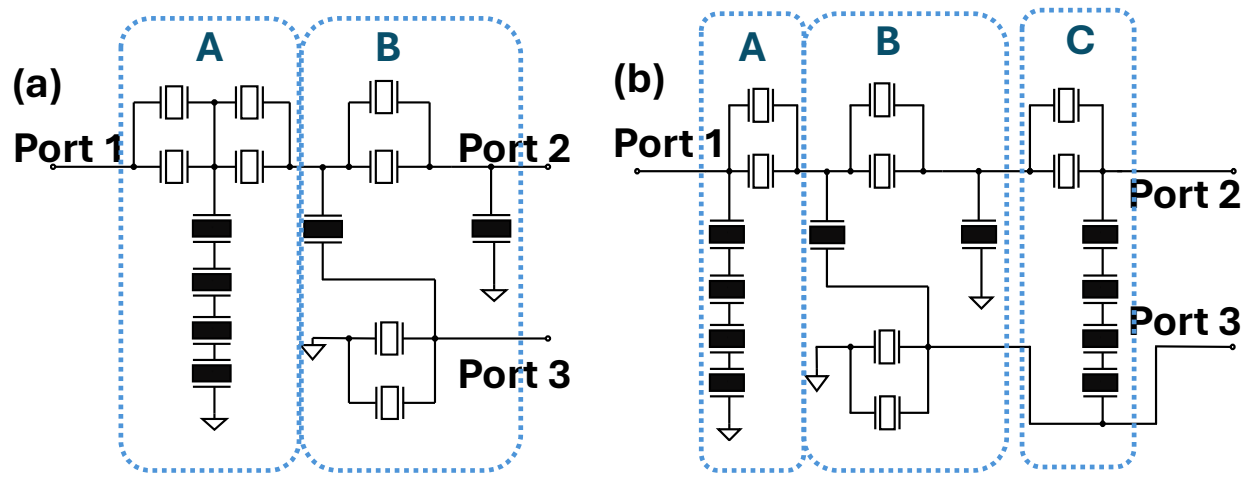

Fig. 2. Schematic of (a) mLL filter design type-1, and (b) mLL filter design type-2. Sections A and C denote ladder units, while B is a modified lattice unit. The series and shunt resonators are indicated by color (black for shunt and white for series). Additionally, the shunt resonators have lower frequencies than the series resonators.

Fig. 3a presents the electrical response of a series and shunt resonator, with key impedance regions or points highlighted. Fig. 3b depicts a simplified version of filter type-1, where resonators in series or in parallel have been reduced to a single equivalent form. The series resonators are denoted by S, while the shunt resonators are denoted by H. All the shunt (H) resonators in both mLL filters were designed to have the same resonance ($f_s$) and anti-resonance ($f_p$) frequencies. Likewise, all the series (S) resonators in the mLL filters were designed to have the same $f_s$ and $f_p$ frequencies. The $f_s$ and $f_p$ of the shunt resonators are lower than those of the series resonators.

Finally, to simplify the discussion, the authors will assume infinite resonator quality factor in this explanation: meaning that the resonators act as short circuits (with zero impedance) at resonance and act as open circuits (with infinite impedance) at anti-resonance.

In regions A, E, and I of Fig. 3a, all the resonators are

capacitive, so filter type-1 acts as a capacitive voltage divider from port 1 to ports 2 and 3. The signals received at ports 2 and 3 are in phase with each other, and thus their difference yields a value much less than the signal at port 2. The equivalent circuit of filter-1 as relates to these three regions is depicted in Fig. 3c I. The voltage-divider behavior of the filter in region E explains the concave dip seen in the passband of its simulated and measured responses. At point B of Fig. 3a, all shunt resonators experience resonance, shunting the input signal to ground and creating a null in the transmission response as depicted in Fig. 3c II. In region C, the shunt resonators are all inductive while the series resonators are capacitive, creating a high-pass filter (HPF) response from port 1 to port 2, but a band-pass filter (BPF) response from port 1 to port 3 (see Fig. 3c III). At points D and F of Fig. 3a, the series resonators are at resonance (or near resonance) and thereby short the input port 1 to port 2, but short port 3 to ground (Fig. 3c IV). The shunt resonators operate at antiresonance at these points and thus provide high impedance (represented as opens in Fig. 3c IV) resulting in a notch between ports 1 and 3. In region G, the series resonators are inductive while the shunt resonators are capacitive, forming a low-pass filter (LPF) response from port 1 to port 2, but a band-pass filter response of opposite phase to that in region C from port 1 to port 3 (see Fig. 3c V). Finally, at point H, the series resonators act as open circuits at their antiresonance frequency, causing a null in the filter's transmission response as the input signal at port 1 is reflected (see Fig. 3c VI). A depiction of the shape of the transmission responses in the discussed regions as viewed from the output ports, as well as the difference of these responses, is shown in Fig. 3d.

With regards to the phase at ports 2 and 3 in regions C and G, note that ports 2 and 3 must be 180° out-of-phase with each other in these regions. Intuitively, we see the opposing LC and CL networks extending from the common node $V_x$ to the output ports, which must be 180° out-of-phase with each other with proper sizing. Mathematically, the output voltages at port 2 ($V_{out2}$) and port 3 ($V_{out3}$), given an input signal from node $V_x$ in Fig. 3c V, are given by

$$V_{out2} = \frac{-1}{-1 + \omega^2 L_S C_H} V_X, \tag{1}$$

and

$$V_{out3} = \frac{\omega^2 L_S C_H}{-1 + \omega^2 L_S C_H} V_X, \tag{2}$$

and are therefore 180° out of phase with each other. Here, $L_S$ denotes the motional inductances of the series resonators, $C_H$ denotes the electrical capacitances of the shunt resonators, $V_x$ is the voltage at node $V_x$, and ω denotes the angular frequency. A similar 180° phase difference is observed in Fig. 3c III, except that the series resonators are capacitive ($C_S$), and the shunt resonators are inductive ($L_H$). The difference between the signals at port 2 and port 3 should thus yield a balanced output signal with amplitude far less than the signal amplitude from port 2 in regions A and I (where the output signals are in phase with each other), but with double the signal amplitude in regions C and G. This relaxes the performance demands on the sub-units of the mLL filter, particularly the ladder section(s).

It is important to note that the quality factors of all the resonators determine the filter's insertion loss (Fig. 3c IV represents the ideal case), and the filter bandwidth is determined by the $k_t^2$ of the resonators – i.e., the separation of their $f_s$ and $f_p$ frequencies (as seen in Fig. 3d). Equally as importantly, the steepness of the passband's skirts is determined by the quality factors of the shunt and series resonators. In Fig. 3c III and V, by designing the cut-off frequencies of the low-pass and high-pass responses in both the ladder and modified lattice segments to be lower than the $f_s$ frequency of the shunt resonators (for the high-pass responses) and greater than the $f_p$ frequency of the series resonators (for the low-pass responses), one can ensure that the steepness of the passband skirt and the roundness of the passband shoulders is determined by how quickly the resonators' phases change from capacitive to inductive or vice-versa in the vicinity of their resonance or anti-resonance frequencies – a change rate that is indicative of their quality factors at resonance/antiresonance. This mechanism can be further understood by observing the transitions from regions A to C or regions G to I as shown in Fig. 3c I, III, and V, which can be accelerated or decelerated based on the quality factors of the resonators at points B, D, F, and H of Fig. 3a. Conventional ladder filters exhibit a similar characteristic [1].

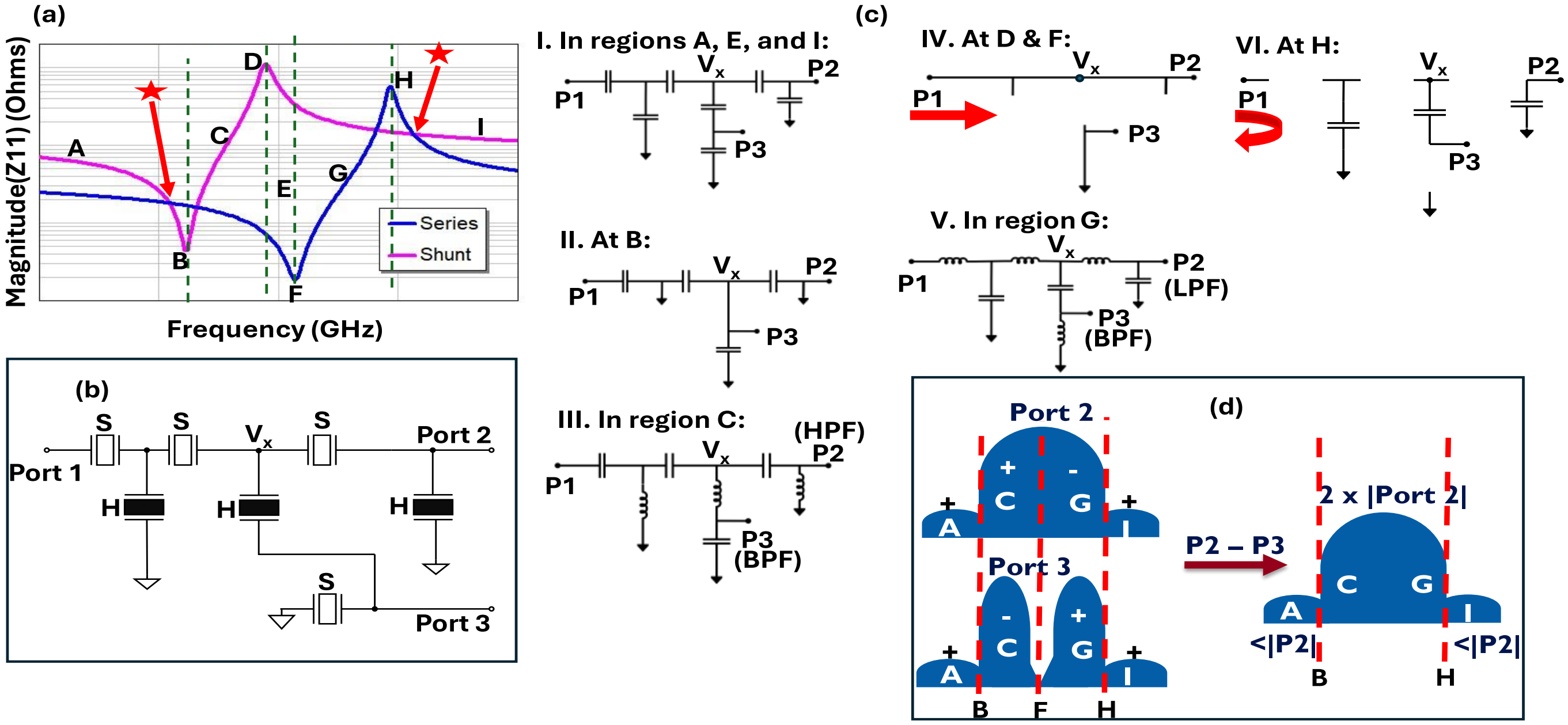

Fig. 3. (a) Electrical impedances of a sample series and shunt resonator, showing key points and regions. (b) The equivalent circuit for filter type-1, (c) representations of the equivalent circuit in (b) at various frequency regions/points, and (d) illustration of the amplitude of the output signals from the balanced port as a function of input frequency.

Like a conventional ladder filter, the mLL filter exhibits transmission zeros at the $f_s$ and $f_p$ frequencies of the shunt and series resonators, respectively. Additionally, like a conventional lattice filter, mLL filter type-1 exhibits transmission zeros in regions A and I when the impedance of the shunt and series resonators are equal. These intersections are indicated by the stars on the impedance plot of Fig. 3a. There, ports 2 and 3 receive equal signals that are in phase with each other and thus yield a zero response when subtracted. These zeros thus occur beyond the resonance frequencies of the resonators, in their capacitive regions, as in conventional balanced lattice filters [29-30], [34]. The mLL filter type-2 also experiences zeros in its transmission response when its output ports 2 and 3 receive equal, in-phase, signals, but this does not occur exactly at the equal-impedance points of its resonators, due to the asymmetry of its output ladder unit (see Fig. 2b).

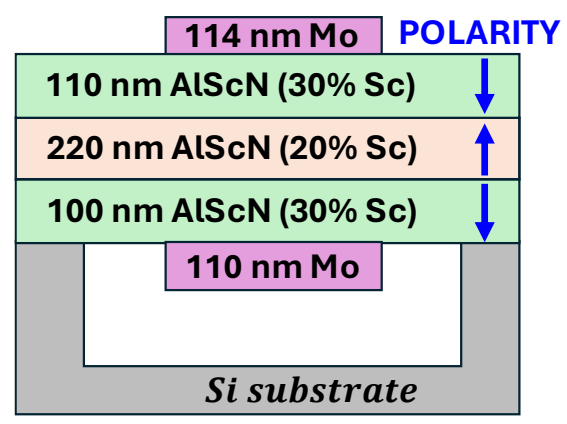

Fig. 4. Material stack of the series resonators in the fabricated mLL filters, with indicated polarity of the P3F AlScN layers.

We conclude that the mLL filter, which incorporates a single-ended ladder and a single-ended to balanced lattice segment, functions identically to a conventional balanced ladder-lattice filter where only the signal through half a mirrored ladder is considered at the terminal end. The modified lattice segment itself functions as a balun, performing LC-CL and CL-LC phase conversions alternatingly (C denotes capacitance, while L denotes inductance). However, the topology presents lower insertion loss than a conventional balanced ladder-lattice topology as it eliminates the signal loss that would occur in the mirror section of balanced ladder segments. It is worth noting that while the series (shunt) resonators of the ladder units had the same frequencies as the series (shunt) resonators of the modified lattice units, they can be made different to realize more complex transmission responses, if permitted by the capabilities of the underlying process technology.

## III. Device Fabrication, Design, and Electromagnetic Simulation

### A. Fabrication

The stack of the series resonators utilized in the mLL filters is shown in Fig. 4. The resonators consist of two polycrystalline $Al_{0.7}Sc_{0.3}N$ layers (with thicknesses of 110 nm and 100 nm) which are deposited nitrogen-polar, and one 220 nm single-crystal $Al_{0.8}Sc_{0.2}N$ layer which is deposited metal-polar [16], [35]. Both the top and bottom electrodes are made of molybdenum, while the interconnects and pads are made from aluminum due to the lower resistivity of aluminum compared to molybdenum. The P3F stacks are fabricated into BAW resonators and filters using the previously detailed P3F Akoustis XBAW$^{TM}$ process [13], [16-17], [20], [22], [35].

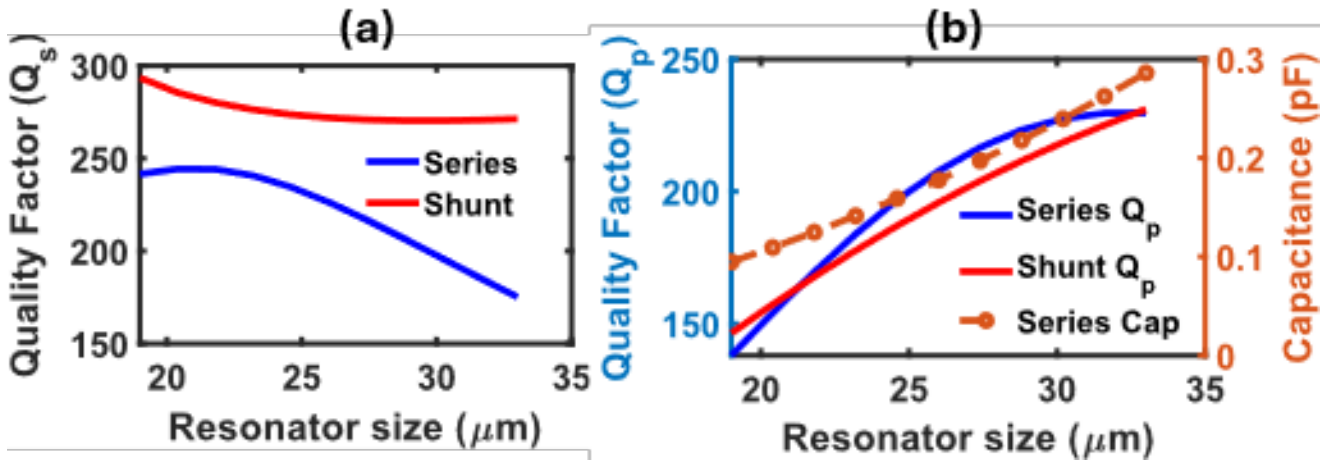

Fig. 5. Quality factors at (a) series resonance frequency ($Q_s$) and (b) antiresonance frequency ($Q_p$) for series and shunt resonators in the design PDK and capacitance ($C_o$) of the series resonators, as a function of resonator size.

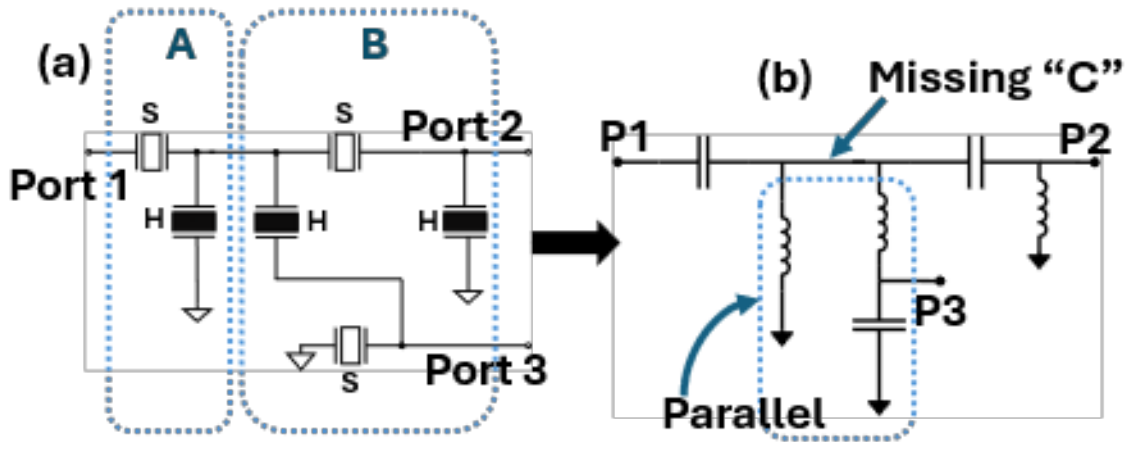

Fig. 6. (a) An alternate mLL design that does not have a series resonator between the ladder "A" and modified lattice "B" units, and (b) its equivalent electrical form at frequencies right above the resonance frequencies ($f_s$) of the shunt resonators. 'S' denotes series resonators and 'H' denotes shunt resonators.

### B. Design

The mLL filters were designed and optimized in the AWR Design Environment (DE) software, using a PDK that contained fixed resonator mBVD models from Akoustis, Inc. The design procedure began with charactering the PDK to determine the available resonance frequencies, electromechanical coupling coefficients, and quality factors. Fig. 5 shows the $Q_s$ and $Q_p$ trends of the series and shunt resonators in the PDK as a function of resonator dimension. The electromechanical coupling coefficients of the PDK's resonators showed little variation with size, remaining between 8.36% and 8.76% for shunt resonators, and between 9.89% and 10.03% for series resonators. As shown in Fig. 5, the $Q_s$ of the resonators decrease with increasing resonator size, while their $Q_p$ increases with increasing resonator size. This indicates a $Q_p$ limited by anchor loss and a $Q_s$ dominated by the electrical series resistance of the interconnects [13], [20], [27].

The design of mLL filter type-1 (Fig. 2a), consisting of a one-and-half-order ladder unit and a fourth-order modified lattice unit, was chosen for the initial design. One order of a ladder unit is defined as a series-shunt resonator pair, while one order of a lattice unit is defined as a single resonator. An additional ladder unit was appended to the chosen topology for filter type-1 to provide higher out-of-band rejection. This design choice led to the design for filter type-2 (Fig. 2b) with a few more modifications. The authors found that connecting the ladder unit to the modified lattice unit through a series resonator is crucial for taking full advantage of the potential of the mLL topology, since it provides a resonator that can form a CL or LC response with the first shunt resonator of the modified lattice unit in the filter passband. The design in Fig. 6a. illustrates this concept. For instance, in the equivalent circuit for Fig. 6a at frequencies just above the shunt $f_s$ (see Fig. 6b), port 2 sees a second-order HPF response since the circled components act in parallel with each other. Adding a connective series resonator would increase the HPF order to three and yield a steeper high-pass response. Therefore, the input segment of an mLL filter should consist of at least one series resonator (a half-order ladder unit).

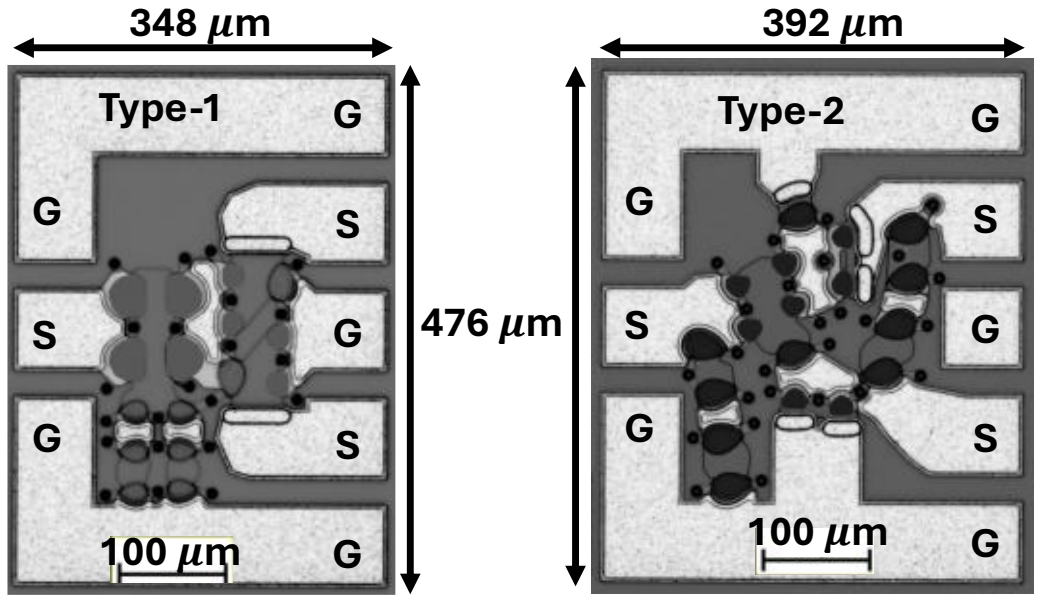

Fig. 7. Micrographs of the fabricated mLL filters. On the left is filter type-1, while filter type-2 is on the right. G denotes the ground pads, while S denotes the signal pads.

Once the topologies were selected, the resonator sizes were optimized using the optimization feature in AWR. The hardest metric to meet was the insertion loss, with a goal of less than 2 dB. Minimum insertion loss favored large shunt resonator sizes and smaller series resonator sizes for the ladder segments, to maximize their $Q_p$ (shunt) and $Q_s$ (series) values as shown by the Q trends in Fig. 5. For mLL filter type-1 which had only two subunits, this pattern could be sacrificed in the ladder subunit for the sake of matching the port impedances. The modified lattice units did the heavy-lifting of supplying out-of-band rejection and hence tended to have similar (effective) resonator sizes for both their shunt and series resonators: since the transmission null caused by the modified lattice units occur at/near the equal impedance point of their constituent resonators. The more closely matched these resonators are in impedance, the higher the out-of-band rejection in the stop bands. In addition to being similar, the modified-lattice resonators tended to be small, as the optimizer prioritized their $Q_s$ and impedance over their $Q_p$, balancing insertion loss and impedance matching to the ports.

When the optimized values of the resonator sizes leaned toward the extremes of the values available in the PDK, the resonators were split into series or parallel combinations whenever meaningful (the ease of layout/ layout symmetry was also considered in making this decision). Fig. 7 shows micrographs of the fabricated filters. Both filters have compact footprints, with filter type-1 occupying a 348 μm x 476 μm area and filter type-2 occupying a 392 μm x 476 μm area. Both layouts were optimized to reduce loss in the interconnects, parasitic capacitances and inductances, as well as mismatches between the balanced output ports. A discussion on the port impedance and electromagnetic simulation of the filters is presented in the supplementary material.

## IV. Measurement and Discussion

### A. S-Parameters

S-parameter measurements were performed on the filters using a 4-port VNA (P5028B from Keysight Technologies), and the setup was calibrated to the tips of the probes. No on-wafer de-embedding was performed. Ports 1 to 3 of the VNA were connected to ports 1 to 3 of the mLL filters, respectively, using GSG-100 probes (I40-A-GSG-100) at the input and GSGSG-100 probes (SP-I67-AM-GSGSG100) at the output. The resulting S-parameter files were then imported into the AWR Design Environment, and their balanced outputs were connected to an ideal balun (component ID: XFMR) to ascertain their differential responses. The authors note that the balun is used only for the sake of characterizing the filters since in the applications for which the filters were designed, they can interface directly with components such as antennas and data converters that have balanced input/output ports [31-33].

Fig. 8 shows the transmission and reflection responses of the fabricated filters, as measured with 50 Ohm differential ports. Filter type-1 operates at a center frequency of 19.03 GHz and achieves 1.29 dB of minimum insertion loss, 6.26% FBW corresponding to a 3-dB bandwidth of 1.91 GHz, and an out-of-band rejection of approximately 30 dB. Filter type-2 operates at a center frequency of 18.82 GHz and achieves a minimum insertion loss of 1.58 dB, a 5.11% FBW corresponding to a 3-dB bandwidth of 961 MHz, and an out-of-band rejection of approximately 33 dB. Both filters display spurious responses, but they are better attenuated in design type-2 than in design type-1. This is in part due to the additional ladder unit in filter type-2 which provides increased rejection of out-of-band signals. The phase and amplitude of transmission responses from the balanced output ports of the filters are shown in Fig. 8c, f, g, and h, and confirm correct operation of the modified ladder-lattice filters. The mLL filter type-1 exhibits zero phase difference between its output ports (defined as the difference between the phase of $S_{21}$ and the phase of $S_{31}$) outside its passband, but experiences phase shifts with greater than 100° absolute values in the passband (see Fig. 8c). Likewise, mLL filter type-2 exhibits a low 20° phase difference outside its passband but sees phase shifts with absolute values greater than 100° within its passband (Fig. 8f). The 20° phase difference is due to the asymmetric ladder unit joining the output ports (2 and 3) of mLL filter type-2 and is observed in simulations as well. The output ports of both filters individually achieve 10- to 15-dB of out-of-band rejection, while the differential responses achieve greater than 30 dB of rejection (Fig. 8 g and h).

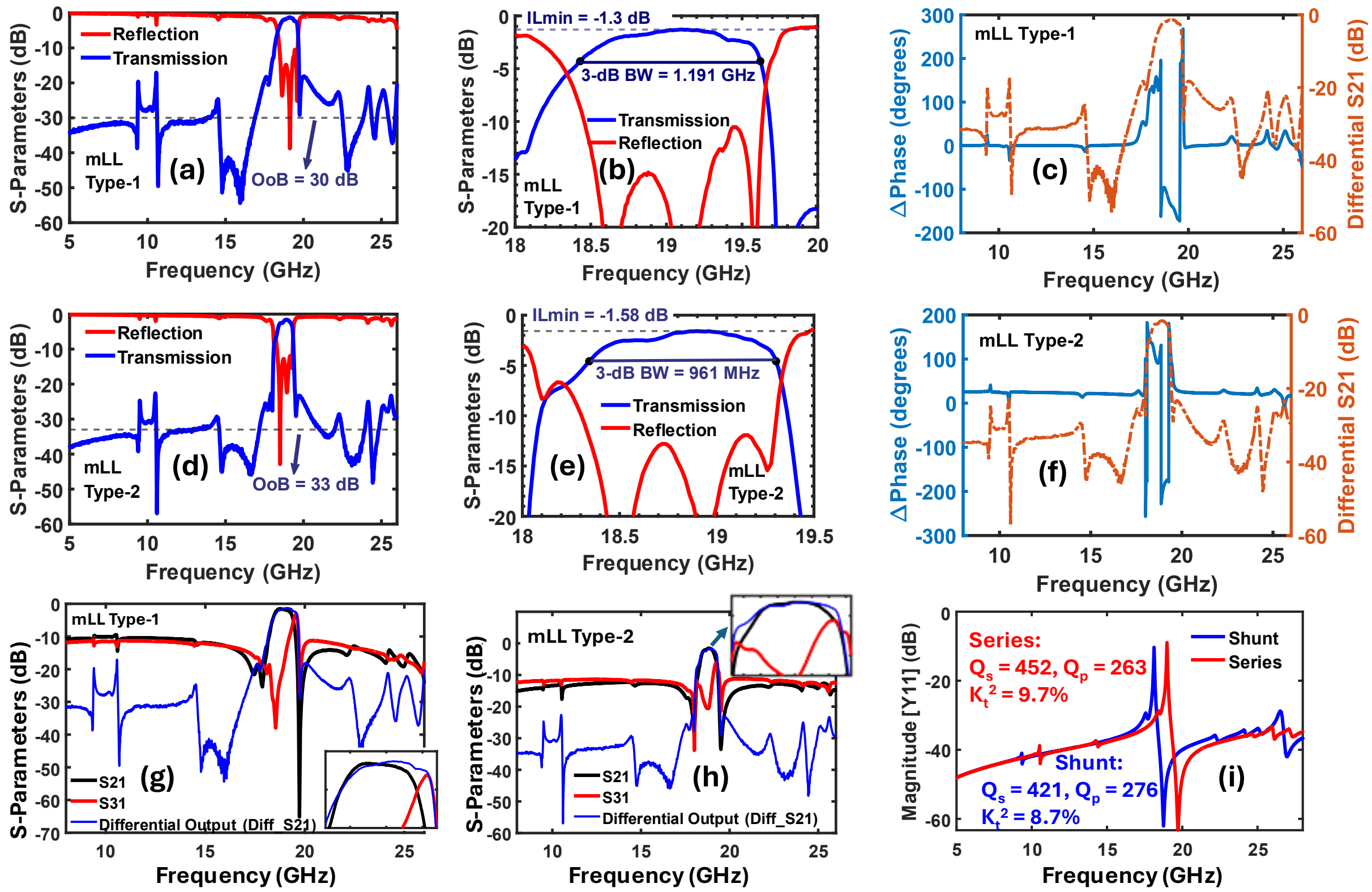

Fig. 8. (a) and (d) show the $S_{21}$ and $S_{11}$ responses of mLL filter type-1 and type-2, respectively. (b) and (e) depict zoom-ins of the filter passbands, showing their bandwidth, insertion loss, and return loss. (c) and (f) show a superposition of the difference in phase between ports 2 and 3 of each filter on their differential response. (g) and (h) show the transmission responses from port 1 to ports 2 and 3, superimposed on the differential response. (i) presents the admittance magnitude of fabricated series and shunt resonators, with indicated resonator metrics. Both resonators have electrical capacitances of 0.127 pF.

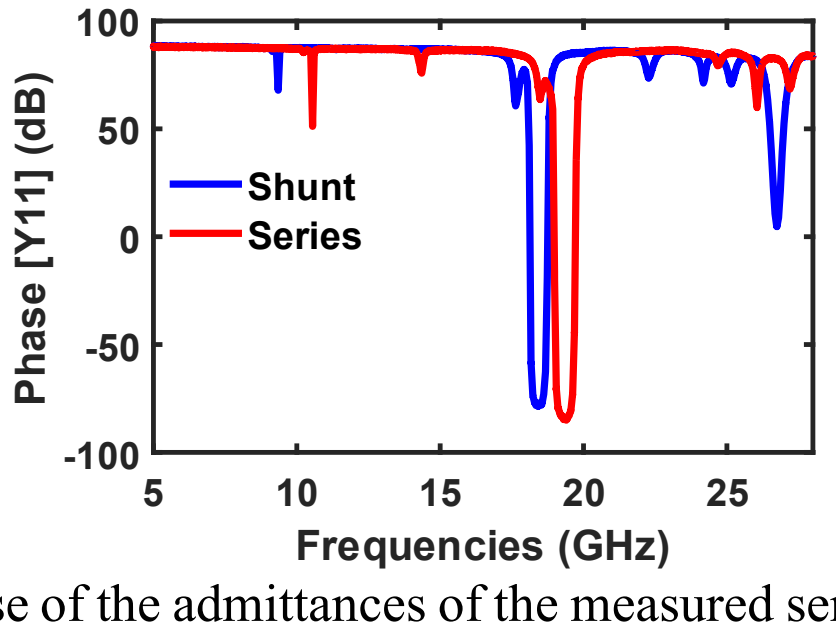

Fig. 9. Phase of the admittances of the measured series and shunt resonators, showing the impact of spurious modes.

The fabricated resonators displayed spurious modes below and near their $f_s$ frequencies as well as above 22 GHz, resulting in corresponding spurious responses in the filters' transmission responses as the phases of the balanced ports are altered from their nominal values (see Fig. 8 c, f, and i). These spurs are due in part to the asymmetric P3F structure (shown in Fig. 4a), which does not perfectly reject lower or higher frequency thickness extensional resonance modes [7], [36]. Fig. 8i displays the admittance of a measured series and shunt resonator, while Fig. 9 shows their phase responses. Finally, a comparison between the lumped simulation, EM-incorporated simulation, and measurements of the mLL filters is shown in Fig. 10. The EM-incorporated simulations better predict the measurement responses compared to the lumped simulations, except for regions containing spurious modes not accounted for in either simulation.

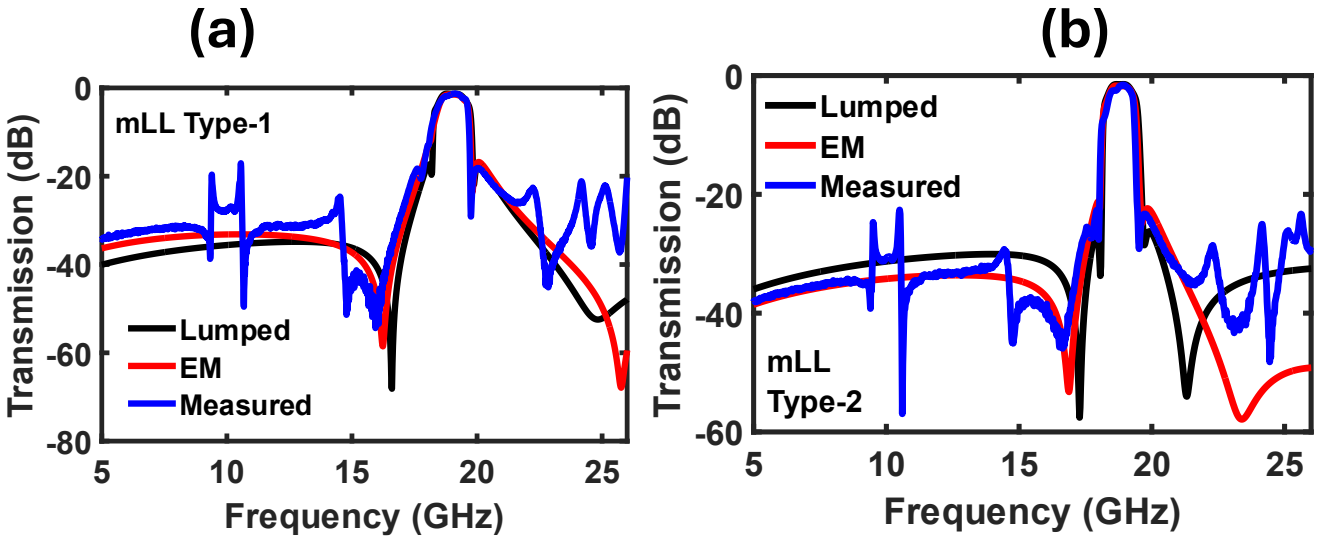

Fig. 10. Comparison of the lumped and EM-incorporated simulations of the dies of (a) filter type-1 and (b) filter type-2 to their measurements, showing a closer agreement between the EM simulation and the measurements in spurious-free regions.

## *B. Linearity*

The third-order intermodulation and power handling of both filter designs were experimentally studied. To ascertain the third-order intermodulation intercept points, power at the tone frequencies of 18.95 GHz and 19.05 GHz (i.e., a 100 MHz tone spacing) were applied at varying amplitudes to the filters, while the output power was measured and detected at the fundamental and intermodulation tones [37]. Both the fundamental tones and the intermodulation products (at 18.85 GHz and 19.15 GHz) fall within the filters' passbands. A 4-port VNA (P5028B from Keysight Technologies) supplied the tones from two ports and measured the output from another port functioning as a spectrum analyzer. Throughout the measurement, port 3 of the device-under-test (DUT) was terminated with a 50-Ohm load (ANNEF-50K+ from Mini-Circuits) while port 2 served as a single-ended output port. As shown in supplementary Fig. S3, mLL filter type-1 had an input third-order intercept point (IIP3) value of 43.52 dBm and output third-order intercept point (OIP3) value of 41.62 dBm, while mLL filter type-2 had an IIP3 value of 43.06 dBm and an OIP3 value of 40.52 dBm. These IP3 values are limited by the measurement setup, whose intermodulation products were determined by measuring a "thru" structure on a 129-239 calibration substrate. The setup's input IP3 was 41.67 dBm, while its OIP3 was 41.56 dBm.

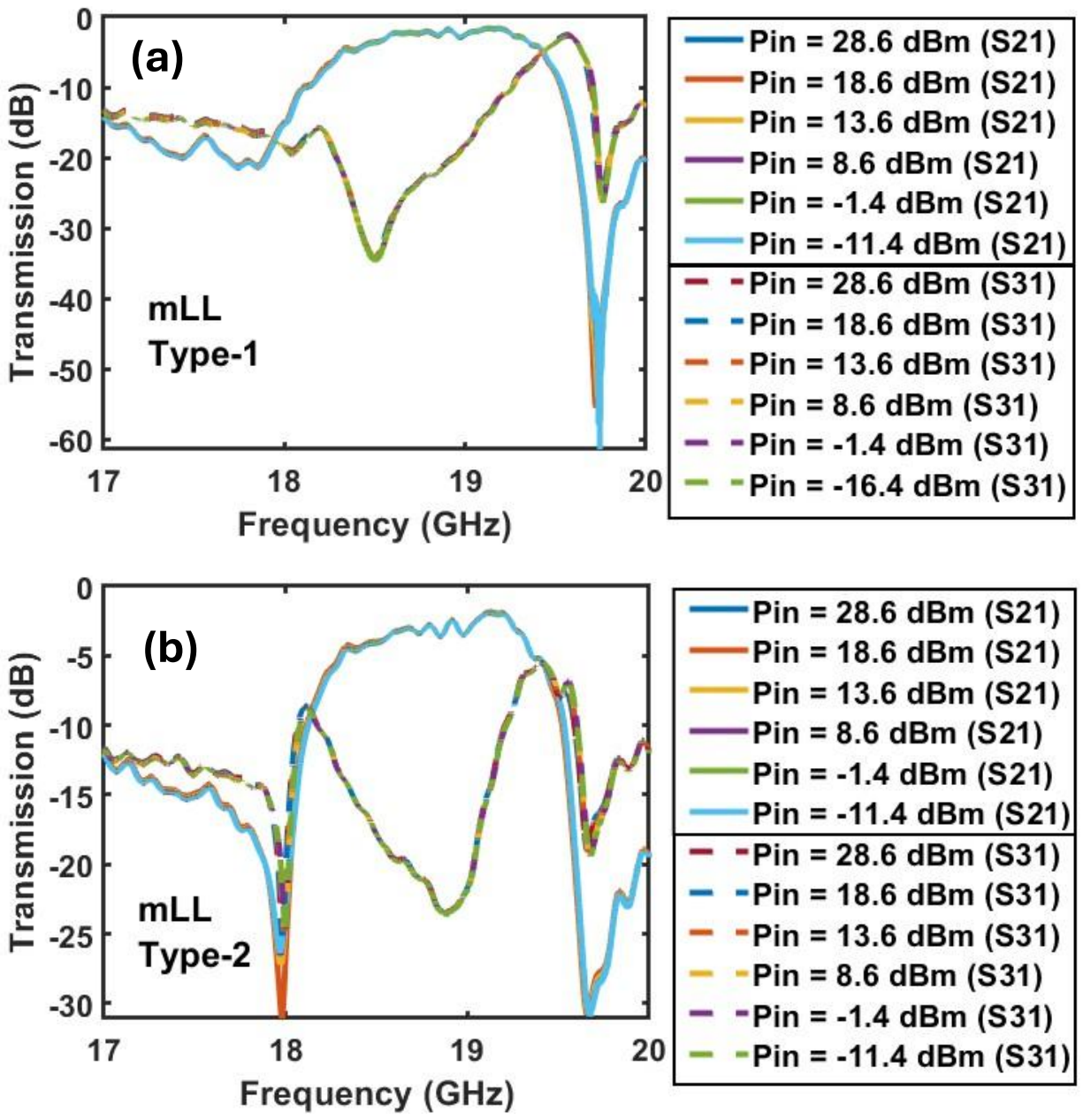

Fig. 11. Power handling measurement with extended input power up to +28.6 dBm for (a) mLL filter type-1 and (b) mLL filter type-2.

Because the 4-port VNA used for the preceding IP3 measurements only supplies a maximum of +8-dBm power at high frequencies, a 2-port VNA (P9374A from Keysight Technologies) with a higher power supply of +20 dBm, but maximum frequency of 20 GHz, was used for power handling measurements. To further increase the input power, the authors placed a 14-dB amplifier (ZVM-273HP+ from Mini-Circuits) between the output port of the VNA and the device under test (DUT), as well as a 10 dB attenuator (BW-S10W5+ from Mini-Circuits) between the DUT and the output-power-detecting port of the VNA to protect the VNA from a power overload. This setup was not calibrated using a SOLT calibration technique. Instead, a transmission (i.e., through) measurement was performed on the thru standard of a 129-239 calibration substrate at varying power levels. This thru response was then subtracted from the measured filter S-parameter responses obtained with the same setup. Additional power handling measurements and setup information can be found in the supplementary material. Fig. 11 shows that the mLL filters can handle input powers up to +28.6 dBm (the limit of the test setup) with gain compression less than 1 dB. The input powers have been calibrated for the loss through the input cables and probes to the probe tips. With the combined high IP3 intercept point and high-power handling, these mLL filters are both miniature and highly linear.

## V. Conclusion

The authors have presented a novel, modified ladder-lattice (mLL) filter topology that can perform single-ended to differential/balanced mode conversion without the use of baluns or passive components. To perform these conversions, the mLL filter employs a modified lattice unit which performs the alternating function of an LC-CL or CL-LC balun within the filter's passband, providing roughly 180$^{o}$ of phase shift between the balanced ports. Though the implementations shown in this paper had relatively low filter orders, more ladder, balanced lattice, and modified lattice units can be appended as needed for specific filtering applications. Apart from its compact size (due to the absence of baluns) and reduced complexity compared to conventional balanced ladder-lattice filters, the mLL filter provides lower insertion loss compared to a fully balanced ladder-lattice filter of the same order, due to the omission of the mirror-sections of the balanced ladder segments. Finally, the frequencies of the constituent resonators in the filter serve as an additional degree of freedom to realize more complex transmission responses.

Two implementations of the mLL filter topology were implemented and fabricated, displaying low insertion loss, moderate bandwidth, high out-of-band rejection, and high linearity. Notably, design type-1 achieved a low minimum insertion loss of 1.3 dB, 6.26% bandwidth, and 30 dB of average out-of-band rejection, while design type-2 achieved 1.58 dB minimum insertion loss, 5.11% fractional bandwidth, and 33-dB of average out-of-band rejection. A major advantage of design-2 over design-1 is its high selectivity, making it more suitable for duplexer applications that require narrow guard bands. Both filters demonstrated IIP3 values greater than 40 dBm and power handling higher than 28 dBm, and occupied footprints less than 400 um x 480 um.

## References

[1] G. Piazza, V. Felmetsger, P. Muralt, R. H. Olsson, and R. Ruby, "Piezoelectric aluminum nitride thin films for microelectromechanical systems," MRS Bulletin, vol. 37, no. 11, pp. 1051–1061, Nov. 2012, doi: 10.1557/mrs.2012.268.

[2] Y. Liu, Y. Cai, Y. Zhang, A. Tovstopyat, S. Liu, and C. Sun, "Materials, Design, and Characteristics of Bulk Acoustic Wave Resonator: A Review," Micromachines (Basel), vol. 11, no. 7, p. 630, Jun. 2020, doi: 10.3390/mi11070630.

[3] R. H. Olsson, Z. Tang, and M. D'Agati, "Doping of Aluminum Nitride and the Impact on Thin Film Piezoelectric and Ferroelectric Device Performance," in 2020 IEEE Custom Integrated Circuits Conference (CICC), Mar. 2020, pp. 1–6. doi: 10.1109/CICC48029.2020.9075911.

[4] G. Fattinger, A. Volatier, R. Aigner, and F. Dumont, "BAW PCS-Duplexer chipset and Duplexer applications," in 2008 IEEE Ultrasonics Symposium, Nov. 2008, pp. 602–606. doi: 10.1109/ULTSYM.2008.0144.

[5] K. Wang, D. Clark, L. H. Camnitz, and P. Bradley, "A UMTS-900 duplexer," in 2008 IEEE Ultrasonics Symposium, Nov. 2008, pp. 899–902. doi: 10.1109/ULTSYM.2008.0217.

[6] M. Moreira, J. Bjurström, I. Katardjev, and V. Yantchev, "Aluminum scandium nitride thin-film bulk acoustic resonators for wide band

applications," Vacuum, vol. 86, no. 1, pp. 23–26, Jul. 2011, doi: 10.1016/j.vacuum.2011.03.026.

[7] W. Dou et al., "Super-High-Frequency bulk acoustic resonators based on aluminum scandium nitride for wideband applications," Nanomaterials, vol. 13, no. 20, p. 2737, Oct. 2023, doi: 10.3390/nano13202737.

[8] "COFFEE: Compact Front-End Filters at the Element-Level | DARPA." Accessed: Feb. 07, 2026. [Online]. Available: https://www.darpa.mil/research/programs/compact-front-end-filters-at-the-element-level.

[9] M. Akiyama, K. Kano, and A. Teshigahara, "Influence of growth temperature and scandium concentration on piezoelectric response of scandium aluminum nitride alloy thin films," Appl. Phys. Lett., vol. 95, no. 16, p. 162107, Oct. 2009, doi: 10.1063/1.3251072.

[10] H. Liu, F. Zeng, G. Tang, and F. Pan, "Enhancement of piezoelectric response of diluted Ta doped AlN," Applied Surface Science, vol. 270, pp. 225–230, Apr. 2013, doi: 10.1016/j.apsusc.2013.01.005.

[11] N. Xiong, J. Wang, H. Yang, B. Ma, and W. Zou, "High-efficiency dual-level heterogenous grating coupler on CMOS-compatible silicon-lithium niobate platform," Appl. Phys. Lett., vol. 125, no. 8, p. 083502, Aug. 2024, doi: 10.1063/5.0215468.

[12] R. Safian, M. Teng, L. Zhuang, and S. Chakravarty, "Foundry-compatible thin film lithium niobate modulator with RF electrodes buried inside the silicon oxide layer of the SOI wafer," Opt. Express, vol. 28, no. 18, p. 25843, Aug. 2020, doi: 10.1364/OE.396335.

[13] Izhar et al., "Periodically poled aluminum scandium nitride bulk acoustic wave resonators and filters for communications in the 6G era," Microsyst Nanoeng, vol. 11, no. 1, p. 19, Jan. 2025, doi: 10.1038/s41378-024-00857-4.

[14] O. Barrera et al., "Thin-Film Lithium Niobate Acoustic Filter at 23.5 GHz With 2.38 dB IL and 18.2% FBW," Journal of Microelectromechanical Systems, vol. 32, no. 6, pp. 622–625, Dec. 2023, doi: 10.1109/JMEMS.2023.3314666.

[15] Z. Schaffer, A. Hassanien, M. A. Masud, and G. Piazza, "A Solidly Mounted 55 GHz Overmoded Bulk Acoustic Resonator," in 2023 IEEE International Ultrasonics Symposium (IUS), Sep. 2023, pp. 1–4. doi: 10.1109/IUS51837.2023.10307494.

[16] R. Vetury et al., "A manufacturable AlScN periodically polarized piezoelectric film bulk acoustic wave resonator (AlScN P3F BAW) operating in overtone mode at X and Ku band," in IEEE MTT-S Int. Microw. Symp. Dig., Jun. 2023, pp. 891–894, doi: 10.1109/IMS37964.2023.10188141.

[17] Izhar et al., "A K-band bulk acoustic wave resonator using periodically poled Al0.72Sc0.28N," IEEE Electron Device Lett., vol. 44, no. 7, pp. 1196–1199, Jul. 2023, doi: 10.1109/LED.2023.3282170.

[18] M. Park, J. Wang, D. Wang, Z. Mi, and A. Ansari, "A 19 GHz allepitaxial Al0.8Sc0.2N cascaded FBAR for RF filtering applications," IEEE Electron Device Lett., vol. 45, no. 7, pp. 1341–1344, Jul. 2024, doi: 10.1109/LED.2024.3404477.

[19] W. Peng, S. Nam, D. Wang, Z. Mi, and A. Mortazawi, "A 56 GHz trilayer AlN/ScAlN/AlN periodically poled FBAR," in IEEE MTT-S Int. Microw. Symp. Dig., Jun. 2024, pp. 150–153, doi: 10.1109/ims40175.2024.10600386.

[20] Izhar et al., "A high quality factor, 19-GHz periodically poled AlScN BAW resonator fabricated in a commercial XBAW process," IEEE Trans. Electron Devices, vol. 71, no. 9, pp. 5630–5637, Sep. 2024, doi: 10.1109/TED.2024.3435175.

[21] S. Rassay, D. Mo, and R. Tabrizian, "Dual-mode scandium-aluminum nitride Lamb-wave resonators using reconfigurable periodic poling," Micromachines, vol. 13, no. 7, p. 1003, Jun. 2022, doi: 10.3390/mi13071003.

[22] M. M. A. Fiagbenu et al., "A High-Selectivity 11.9-GHz Filter Realized in a Periodically Poled Piezoelectric AlScN Film," IEEE Microwave and Wireless Technology Letters, pp. 1–4, 2025, doi: 10.1109/LMWT.2025.3581841.

[23] O. Barrera et al., "Transferred thin film lithium niobate as millimeter wave acoustic filter platforms," in Proc. IEEE 37th Int. Conf. Micro Electro Mech. Syst. (MEMS), Jan. 2024, pp. 23–26, doi: 10.1109/mems58180.2024.10439593.

[24] S. Cho et al., "23.8-GHz acoustic filter in periodically poled piezoelectric film lithium niobate with 1.52-dB IL and 19.4% FBW," IEEE Microw. Wireless Technol. Lett., vol. 34, no. 4, pp. 391–394, Apr. 2024, doi: 10.1109/LMWT.2024.3368354.

[25] X. Tong et al., "6 GHz Lamb wave acoustic filters based on A1-mode lithium niobate thin film resonators with checker-shaped electrodes," Microsyst. Nanoeng., vol. 10, no. 1, pp. 1–10, Sep. 2024, doi: 10.1038/s41378-024-00776-4.

[26] O. Barrera, S. Cho, J. Kramer, V. Chulukhadze, J. Campbell, and R. Lu, "38.7 GHz thin film lithium niobate acoustic filter," in Proc. IEEE Int. Microw. Filter Workshop (IMFW), Feb. 2024, pp. 87–90, doi: 10.1109/imfw59690.2024.10477121.

[27] C. Muller and M.-A. Dubois, "Effect of size and shape on the performances of BAW resonators: A model and its applications," in Proc. IEEE Ultrason. Symp., Nov. 2008, pp. 1552–1556, doi: 10.1109/ULTSYM.2008.0378.

[28] Q. Yang, W. Pang, D. Zhang, and H. Zhang, "A Modified Lattice Configuration Design for Compact Wideband Bulk Acoustic Wave Filter Applications," Micromachines (Basel), vol. 7, no. 8, p. 133, Aug. 2016, doi: 10.3390/mi7080133.

[29] A. A. Shirakawa et al., "Ladder-lattice bulk acoustic wave filters: Concepts, design, and implementation," International Journal of RF and Microwave Computer-Aided Engineering, vol. 18, no. 5, pp. 476–484, 2008, doi: 10.1002/mmce.20306.

[30] H. K. J. ten Dolle, J. W. Lobeek, A. Tuinhout, and J. Foekema, "Balanced lattice-ladder bandpass filter in bulk acoustic wave technology," in 2004 IEEE MTT-S International Microwave Symposium Digest, Jun. 2004, pp. 391-394 Vol.1. doi: 10.1109/MWSYM.2004.1335904.

[31] H. Zhang and W. Pang, "A Novel Single-Ended to Balanced Bulk Acoustic Wave Filter for Wireless Communications," IEEE Microwave and Wireless Components Letters, vol. 21, no. 7, pp. 347–349, Jul. 2011, doi: 10.1109/LMWC.2011.2145415.

[32] S. Seshadri, K. Tejani, and C. Ritchey, "Advantages of using differential to single-ended RF amplifiers in a transmit signal chain design," 2024.

[33] Y. Zhang, "Differential Antennas: Fundamentals and Applications," Electromagnetic Science, vol. 1, no. 1, pp. 1–17, Mar. 2023, doi: 10.23919/emsci.2022.0002.

[34] Akoustis Technologies. A10252. Accessed: May 16, 2025. [Online]. Available: https://akoustis.com/product/a10252/.

[35] C. G. Moe, J. Leathersich, D. Carlstrom, F. Bi, D. Kim, and J. B. Shealy, "Metal–organic chemical vapor deposition-grown AlScN for microelectromechanical-based acoustic filter applications," Phys. Status Solidi (A), vol. 220, no. 16, Aug. 2023, Art. no. 2200849, doi: 10.1002/pssa.202200849.

[36] W. Pang, H. Zhang, H. Yu, and E. S. Kim, "Analytical and experimental study on second harmonic response of FBAR for oscillator applications above 2GHz," in IEEE Ultrasonics Symposium, 2005., Sep. 2005, pp. 2136–2139. doi: 10.1109/ULTSYM.2005.1603304.

[37] X. Du et al., "Frequency tunable magnetostatic wave filters with zero static power magnetic biasing circuitry," Nature Commun., vol. 15, no. 1, p. 3582, Apr. 2024, doi: 10.1038/s41467-024-47822-3.

**Authors'** photographs and biographies are not available at the time of publication.

## Supplementary Information

# 19 GHz Single-Ended-to-Balanced Modified Ladder-Lattice Filters Realized Using Periodically Polarized AlScN BAW Resonators

Merrilyn M. A. Fiagbenu, *Student Member, IEEE*, Shun Yao, *Graduate Student Member, IEEE*, Siddhant Sahoo, Mojtaba Hodjat-Shamami, *Member, IEEE*, Daeho Kim, *Member, IEEE*, Craig Moe, *Member, IEEE*, Pinal Patel, Ramakrishna Vetury, and Roy H. Olsson, III, *Senior Member, IEEE*

### Supplementary Note 1: Port Impedance

The port impedances utilized for the simulations were 50 Ohm at the input port and 50 Ohm at the output port. An ideal balun (component ID: XFMR) was used in AWR to view the differential response. Fig. S1a shows a simplified depiction of the port connections. The output port impedance can be modified to suit the requirements of the desired application. For instance, for impedance matching to typical 50 Ohm systems, a 100 Ohm termination at the output of Fig. S1a can be used to model the impedance seen looking into the balanced ports of a differential 50-Ohm amplifier or differential 25-Ohm antenna, for instance [1]. A comparison of simulations of the transmission response of the mLL filter type-1 with a 50 Ohm versus a 100 Ohm balanced output port is shown in Fig. S1b for comparison. The out-of-band responses remain roughly the same, but the insertion loss increases by roughly 0.7 dB at the edge of the passband. This makes sense since this implementation was optimized for a 50 Ohm balanced port. Unless stated otherwise, all presented differential responses were determined with 50 Ohm differential port impedances.

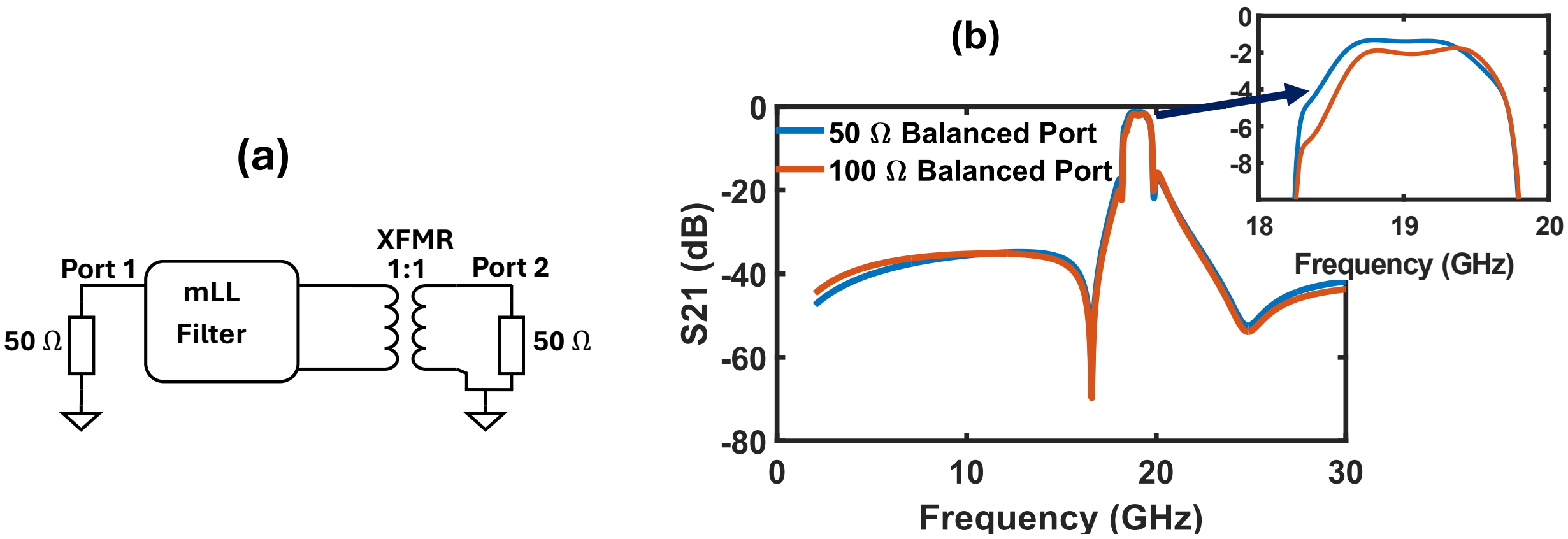

Fig. S1. (a) A depiction of the port connections during circuit simulations in the AWR Design Environment. (b) Comparison of the transmission responses for filter type-1 when simulated with a 50 Ohm vs. a 100 Ohm balanced port. The insert is a zoom-in of the passbands.

## Supplementary Note 2: Electromagnetic (EM) Simulation

To better predict the response of the fabricated filters, electromagnetic simulation of the filter dies was performed in ANSYS HFSS (High-Frequency Structure Simulator). The benefits of such a simulation include better prediction of parasitic capacitances, inductances, losses, and electromagnetic coupling in the dies [2], [3], [4]. The HFSS EM simulation yields a multi-port S-parameter file that can be imported into a circuit simulator for further analysis.

Ground-Signal-Ground (GSG) and Ground-Signal-Ground-Signal-Ground (GSGSG) input and output probes were modeled with lumped port models, using the port structure described by Johansen et al. (see Fig. S2) [5]. However, unlike Johansen et al., the authors did not perform L-2L calibration (or any other calibration) on the simulation results [5]. The resonators were modeled as lumped ports placed between top and bottom electrodes in the HFSS models of the filter dies [2], [3]. A custom material with zero relative permittivity was placed between the top and bottom electrodes of each resonator to yield zero capacitance, so that the full mBVD model of the resonators could be subsequently connected to the output S-parameter file from HFSS without any modifications. This is different from the approaches used by Fattinger et al. or Yang et al., where they account for the static capacitance of the resonators in the EM model and then connect only the acoustic branch of their respective BVD models to the output S-parameter file from HFSS in a linear circuit simulator [2], [3]. This work's EM modeling approach thus sacrifices some accuracy for a simplified design process since the resonator models are fixed in the PDK and cannot be changed. The circuit simulator used in this work for simulating the EM-incorporated response is again AWR DE.

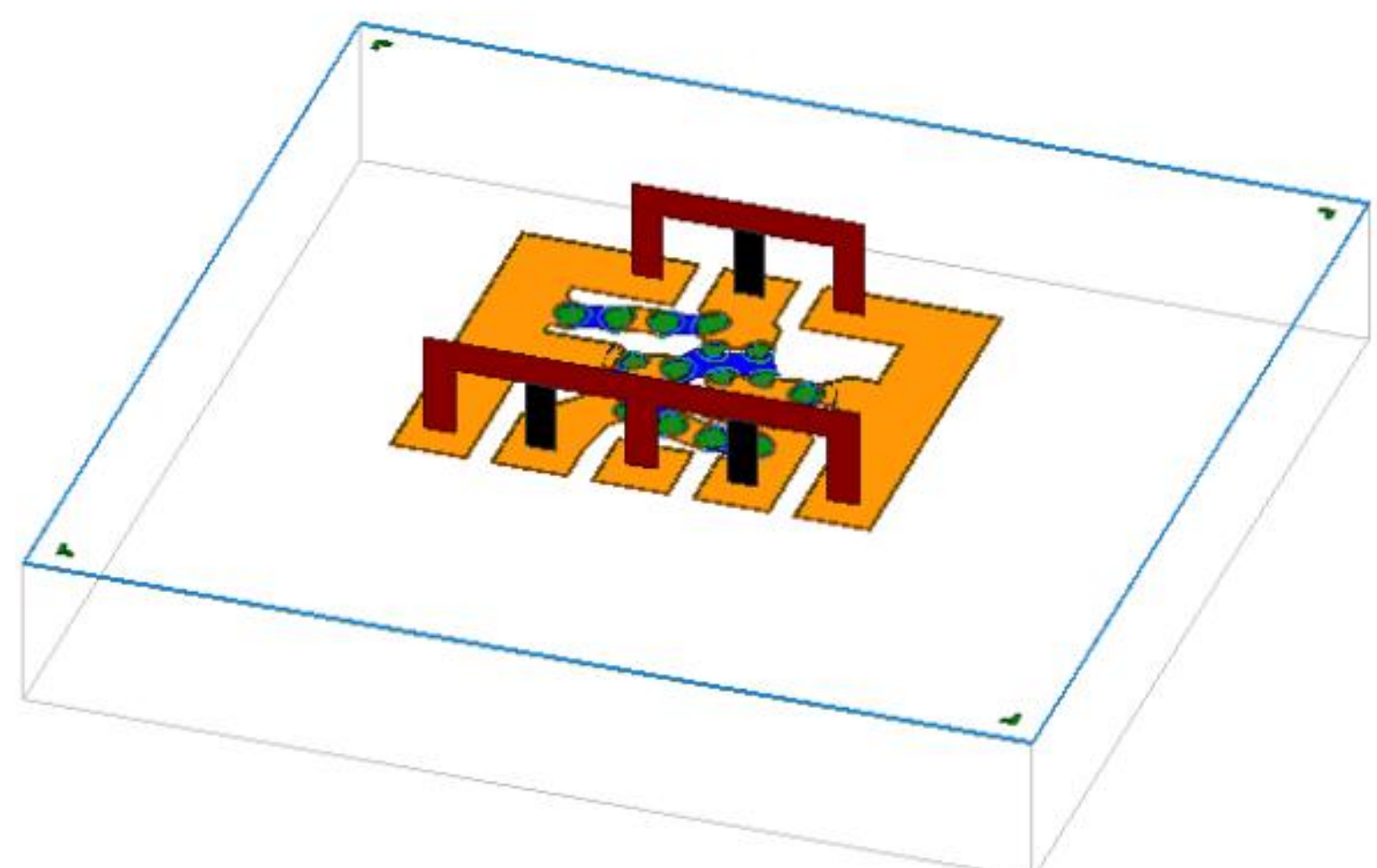

Fig. S2. EM model of filter type-2 in ANSYS HFSS, showing the filter die, the black input and output lumped port models, and the brown PEC ground bridges.

**Supplementary Note 3: IIP3 Measurement**

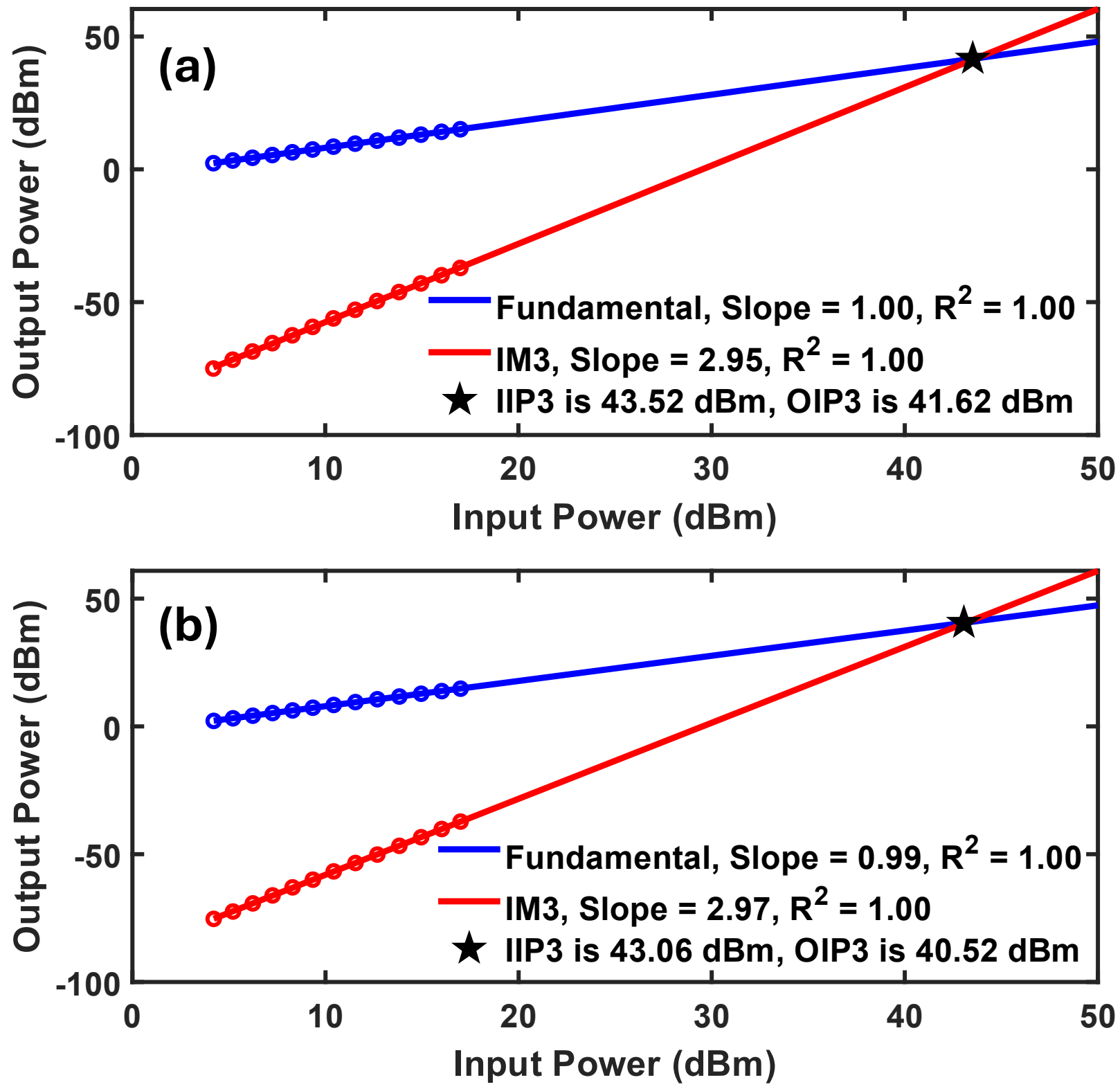

Fig. S3. IP3 measurements of (a) mLL filter type-1 and (b) mLL filter type-2, showing high linearity of both filters. The IP3 in both cases is limited by the measurement setup.

**Supplementary Note 4: Power Handling Measurement**

A 2-port VNA (P9374A from Keysight Technologies) with a high power supply of +20 dBm, but maximum frequency of 20 GHz, was used for power handling measurements. Since the mLL filters are three-port devices, their $S_{21}$ responses were measured while their port 3 were terminated with 50-Ohm loads (ANNEF-50K+ from Mini-Circuits), and their $S_{31}$ responses were measured while their port 2 were terminated with similar 50-Ohm loads. Since the 1-dB compression point of the filters could be determined from the transmission responses on both ports, the authors did not measure the $S_{23}$ responses (though this would have enabled the determination of the balanced output response). To obtain the $S_{23}$ response of an mLL filter, however, terminating port 1 with a 50-Ohm load, while using port 2 as a single-ended input and port 3 as a single-ended output would be the approach. The setups were calibrated to the tips of the probes using the SOLT technique. Accounting for the 2.7 dB power loss through the input cable and probe, no changes were observed in the filters' transmission responses even up to a maximum incident power of +17.3 dBm. Fig. S4 shows the power handling measurement for mLL filter type-1.

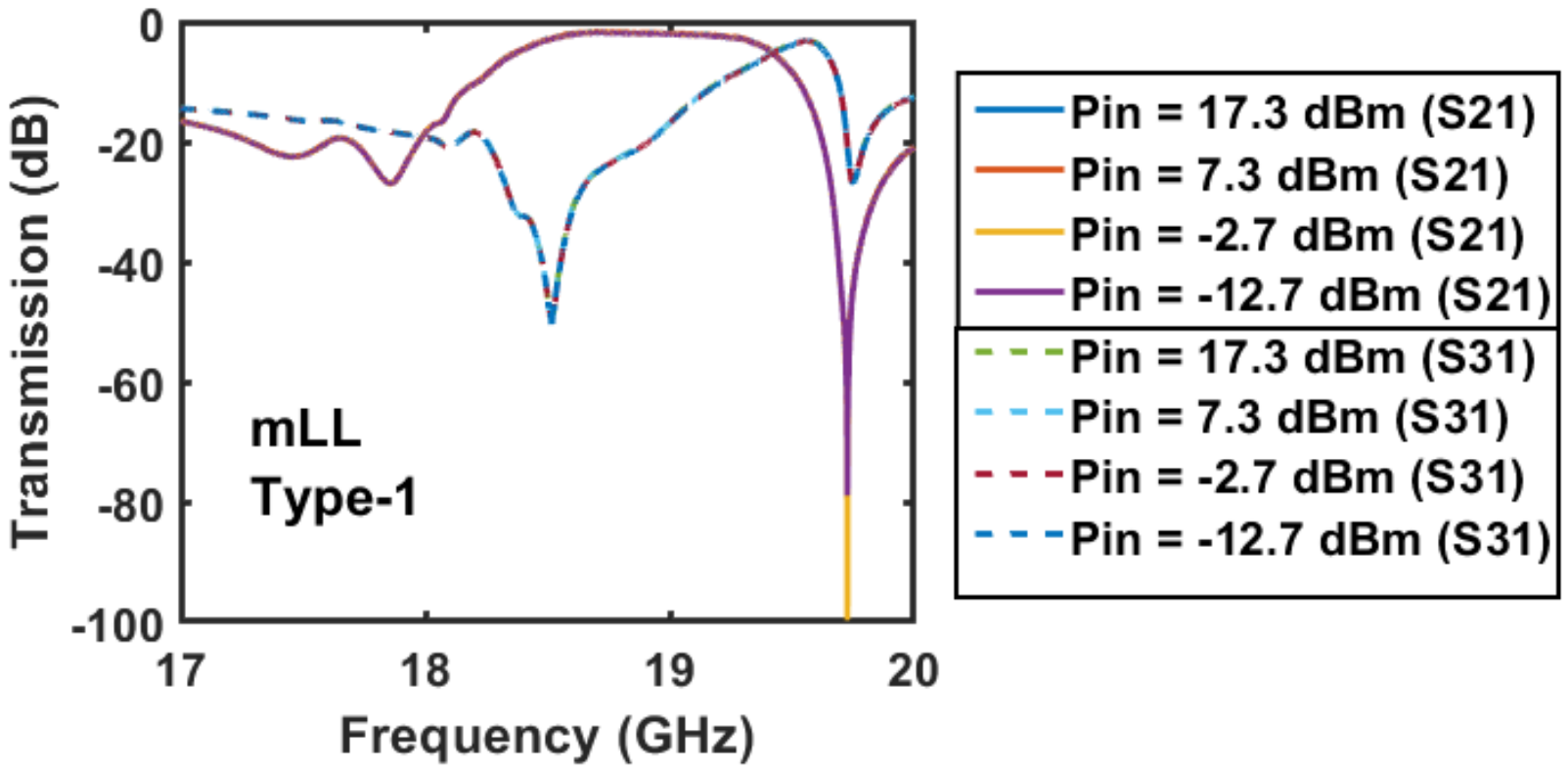

Fig. S4. Power handling measurements on mLL filter type-1 up to +17.3 dBm input power showing no changes in the transmission responses from port 1 to port 2 or port 3.

**References**

[1] R. F. Milsom, H.-P. Lobl, D. N. Peligrad, J.-W. Lobeek, A. Tuinhout, and R. H. ten Dolle, “Combined acoustic-electromagnetic simulation of thin-film bulk acoustic wave filters,” in 2002 IEEE Ultrasonics Symposium, 2002. Proceedings., Oct. 2002, pp. 989–994 vol.1. doi: 10.1109/ULTSYM.2002.1193562.

[2] G. Fattinger, A. Volatier, R. Aigner, and F. Dumont, “BAW PCS-Duplexer chipset and Duplexer applications,” in 2008 IEEE Ultrasonics Symposium, Nov. 2008, pp. 602–606. doi: 10.1109/ULTSYM.2008.0144.

[3] Q. Yang, W. Pang, D. Zhang, and H. Zhang, “A Modified Lattice Configuration Design for Compact Wideband Bulk Acoustic Wave Filter Applications,” Micromachines (Basel), vol. 7, no. 8, p. 133, Aug. 2016, doi: 10.3390/mi7080133.

[4] R. F. Milsom, H.-P. Lobl, D. N. Peligrad, J.-W. Lobeek, A. Tuinhout, and R. H. ten Dolle, “Combined acoustic-electromagnetic simulation of thin-film bulk acoustic wave filters,” in 2002 IEEE Ultrasonics Symposium, 2002. Proceedings., Oct. 2002, pp. 989–994 vol.1. doi: 10.1109/ULTSYM.2002.1193562.

[5] T. K. Johansen, C. Jiang, D. Hadziabdic, and V. Krozer, “EM simulation accuracy enhancement for broadband modeling of on-wafer passive components,” in Proc. Eur. Microw. Integr. Circuit Conf., Oct. 2007, pp. 447–450, doi: 10.1109/EMICC.2007.4412745.